\documentclass[%
 reprint,
superscriptaddress,
 amsmath,amssymb,
 aps,
twocolumn]{revtex4-2}

\usepackage{graphicx}
\usepackage{dcolumn}
\usepackage{bm}
\usepackage[dvipsnames]{xcolor}
\usepackage{stackrel}

\usepackage{amsfonts}

\usepackage{bbold}

\usepackage{hyperref}
\usepackage[shortlabels]{enumitem}

\usepackage{multirow}
\newcommand{\on}{q_{\rm on}}
\newcommand{\off}{q_{\rm off}}
\usepackage{lineno}

\AtBeginDocument{%
  \setlength{\linenumbersep}{\dimexpr\oddsidemargin+0.48in-\linenumberwidth\relax}%
}

\begin{document}

\preprint{APS/123-QED}


\title{Mass changes the diffusion coefficient of particles with ligand-receptor contacts \\ in the overdamped limit}

\author{Sophie Marbach}
\affiliation{%
 Courant Institute of Mathematical Sciences, New York University,
NY, 10012, U.S.A.
}%
\affiliation{CNRS, Sorbonne Universit\'{e}, Physicochimie des Electrolytes et Nanosyst\`{e}mes Interfaciaux, F-75005 Paris, France }
\email{sophie@marbach.fr, holmes@cims.nyu.edu}
\author{Miranda Holmes-Cerfon}%
\affiliation{%
 Courant Institute of Mathematical Sciences, New York University, 
NY, 10012, U.S.A.
}%


\begin{abstract} 
Inertia does not generally affect the long-time diffusion of passive overdamped particles in fluids. 
Yet a model starting from the Langevin equation predicts a surprising property of particles coated with ligands, that bind reversibly to surface receptors -- 
 heavy particles diffuse more slowly than light ones of the same size. We show this by simulation and by deriving an analytic formula for the mass-dependent diffusion coefficient in the overdamped limit. We estimate the magnitude of this effect for a range of biophysical ligand-receptor systems, and find it is potentially observable for tailored micronscale DNA-coated colloids.

\end{abstract}

\maketitle

It is well known that inertia does not affect either the equilibrium probabilities  or dynamics in overdamped systems \cite{Cates:2015ik}, and especially that it does not affect the long time single-particle diffusion coefficient of micron-scale particles in liquids at equilibrium \cite{bian2016111}. Momentum relaxation for micronscale particles occurs over a timescale $\tau_m \simeq 1~\mathrm{\mu s}$~\footnote{See Supplementary \S3.1}, while
particles are generally observed on much longer timescales, where the equilibrium motion is diffusive with diffusion coefficient independent of mass for large enough particles~\cite{schmidt2003hydrodynamic,balboa2013stokes}. 
Inertia can only affect the short-time mobility of a particle \cite{bian2016111,balboa2013stokes}, and it can play a role for active particles where $\tau_m$ is comparable to the diffusive rotational timescale, which is experimentally accessible only in air~\cite{lowen2020inertial}.
To our knowledge, there is currently no proposed physical system  where inertia could affect the long time single-particle diffusion of micronscale particles in liquids at equilibrium. 



\begin{figure}[h!]
    \centering
    \includegraphics{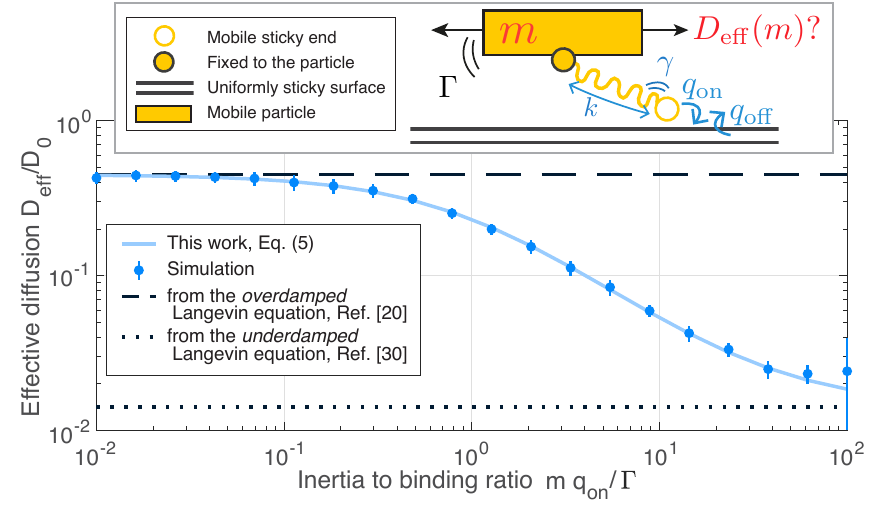}
    \caption{\textbf{Mass changes the diffusion coefficient of a particle with 1 ligand}. (Top) Sketch of a particle of mass $m$ 
    with a single ligand (leg); other parameters described in text. (Bottom) Long-time diffusion coefficient $D_{\rm eff}$ 
    of the particle as a function of inertia to binding timescale ratio $m\on/\Gamma$, obtained from stochastic simulations of Eq.~(\ref{eq:unbound}-\ref{eq:constitutiveBound}) and compared with analytic results, as described in legend. 
    Error bars are one standard deviation for 20 independent simulations, with $\on = 0.01 k/\Gamma$, $\off = 0.008 k /\Gamma$, $\gamma = \Gamma$.}
    \label{fig:intro}
\end{figure}

Yet, the overdamped dynamics of particles with ligand-receptor contacts, such as colloids functionalized with DeoxyriboNucleic Acid (DNA)~\cite{macfarlane2011nanoparticle,rogers2016using,lewis2020single}, viruses~\cite{mammen1998polyvalent,sakai2017influenza,sakai2018unique,muller2019mobility} or white blood cells~\cite{alon2002rolling,ley2007getting,korn2008dynamic}, are not fully understood. 
 Such particles are coated with sticky ligands that bind and unbind to receptors on an opposing surface, 
 changing the particle's mobility~\cite{mammen1998polyvalent,bressloff2013stochastic,hammer2014adhesive,rogers2016using}.
The ligand binding and unbinding rates can be  fast, 
in some cases comparable to $1/\tau_m$ ~\cite{schwarz2004selectin,marbach2021nano}. 
One might speculate that when binding occurs on the same timescale as the relaxation of the ambient fluid's momentum, 
the coupling between binding dynamics and momentum relaxation could lead to inertial effects at longer timescales~\cite{schmidt2003hydrodynamic,balboa2013stokes,bian2016111,li2013brownian,daddi2016long,trulsson2012transition,boyer2011unifying,demery2015mean,lowen2020inertial,dai2020phase,lindner1999inertia}.
For example, bimolecular reactants with inertia can show different survival probability decay functions depending on their mass~\cite{dorsaz2010inertial,piazza2013irreversible}. Furthermore, we have recently pointed out that models of DNA-coated colloids find different long-time diffusion coefficients when they start with the underdamped Langevin equation for particle motion \cite{lee2018modeling} (Fig.~\ref{fig:intro}, dotted line) or from the overdamped equation~\cite{marbach2021nano} (Fig.~\ref{fig:intro}, dashed line). 
We therefore ask: could inertia affect the long-time diffusion of particles with  ligand-receptor contacts? 

Here, we shift the common perspective on overdamped systems by showing that the long-time diffusion coefficient of particles with ligand-receptor contacts depends on mass. We investigate a minimal model for such particles that includes the essential ingredients of (a) inertial relaxation and (b) stochastic dynamics of binding and unbinding. 
We consider an $N$-legged particle of mass $m$, with spring-like legs representing the ligands~\cite{rubinstein2003polymer,miller2006mechanical,lim2006flexible}, and a sticker at the tip of each leg that may transiently attach to a uniformly sticky surface (Fig.~\ref{fig:intro}, inset). Using standard coarse-graining techniques~\cite{pavliotis2008multiscale,fogelson2018enhanced,marbach2021nano}, we derive an analytic expression, verified by simulations, that shows the long-time diffusion coefficient $D_{\rm eff}(m)$ decreases with mass (Fig.~\ref{fig:intro}, blue line). 
This inertial slow-down occurs when the binding timescale is comparable to the inertial timescale, with a magnitude increasing with leg stiffness and decreasing with $N$, and we show it could potentially be measured in DNA-coated colloids with targeted experiments.

Our model starts with the approach we introduced for overdamped dynamics in Ref.~\cite{marbach2021nano} as we have shown it reproduces the experimentally observed diffusion of certain DNA-coated colloids, yet we modify the model to include a dependence on mass. 
The particle's motion is investigated in 1D, on the lateral dimension parallel to the surface. Legs attach and detach independently to the surface with constant rates $\on$, $\off$.  
When unbound, the leg lengths $l_j$ evolve according to 
\begin{equation}
    \frac{dl_j}{dt} =  - \frac{k}{\gamma} (l_j-l_0) + \sqrt{\frac{2 k_B T}{\gamma}} \eta_j\,,
     \label{eq:unbound}
\end{equation}
where $k$ is a spring constant~\cite{rubinstein2003polymer,miller2006mechanical,lim2006flexible}, 
$l_0$ is a rest length and $\gamma$ is the friction coefficient of each leg. The $\eta_j$ are uncorrelated white gaussian noises, such that $\langle \eta_j(t) \eta_k(t) \rangle = \delta_{kj} \delta(t-t')$ and $\langle \eta_j(t)\rangle = 0$ where $\langle \cdot \rangle$ is an average over realizations of the noise.
Inertia of the legs may be neglected as in general legs are much lighter than the particle (Supplementary \S2.4). 
When bound, the legs are constrained to move at the same speed as the particle, $v=dx/dt$, where $x$ is the particle position:
\begin{equation}
       \frac{dx}{dt} = \frac{dl_i}{dt} = v.
       \label{eq:dx}
\end{equation}
Finally, the particle's velocity is governed by Newton's law, including friction $\Gamma$ and stochastic forces $\sqrt{2 k_B T \Gamma} \eta_x$ induced by the ambient fluid, as well as friction, recoil and stochastic forces originating from the bound legs:
\begin{equation}
\begin{split}
       m \frac{dv}{dt} &= - \Gamma v + \sqrt{2 k_B T \Gamma} \eta_x  \\
       &+  \sum_{i\in \, \text{bound}} \left( - \gamma v - k (l_i - l_0) + \sqrt{2 k_B T \gamma} \eta_i \right).
       \end{split}
    \label{eq:constitutiveBound}
\end{equation}
In the absence of legs, the particle diffuses with a bare diffusion coefficient $D_0 = k_B T /\Gamma$. 
The hydrodynamic friction coefficient $\Gamma$ depends on the distance to the wall and may be obtained from lubrication theory \cite{goldman1967slow,sprinkle2020driven} or from measurements \cite{perry2015two}. 
Here the $\eta_i$ and $\eta_x$ are further uncorrelated white Gaussian noises and $i$ is a running index over currently bound legs. 
For a particle in a fluid the relevant mass scale in Eq.~\eqref{eq:constitutiveBound} is $m \rightarrow m + m_f/2$ where $m_f$ is the mass of the displaced volume of fluid~\cite{usabiaga2014inertial,bian2016111}. We remark that in contrast to previous models~\cite{marbach2021nano,holmes2016stochastic}, 
here it is not necessary to project the unbound stochastic dynamics to obtain the bound dynamics; Newton's law is sufficient. 
All parameters of the model, including the binding rates, may depend on temperature $T$, which we assume to be constant. 

The Langevin dynamics  in Eq.~\eqref{eq:constitutiveBound} are a common starting point to investigate the effect of inertia~\cite{demery2015mean,bae2021inertial}. 
Although these equations imply an exponential decay of momentum, which is faster than the algebraic decay that occurs in fluids of similar density as the particle, as envisioned here~\cite{bian2016111}, we expect our model will give a lower bound on the effect of inertia, and is therefore suited to explore the onset of inertial effects. 



Stochastic simulations of our model show that the long-time diffusion coefficient $D_{\rm eff}$ depends on particle mass (see simulation details in Supplementary \S1, building on Ref.~\cite{vanden2006second}). For example,  Fig.~\ref{fig:intro} shows that $D_{\rm eff}$ 
for a 1-legged particle 
 continuously decreases with mass, by more than an order of magnitude, from the overdamped~\cite{marbach2021nano} to the underdamped~\cite{lee2018modeling} regimes. 

To further understand how this decrease depends on model parameters $(m, N, k, \on, \off, \gamma, \Gamma)$, we derive an analytic expression for $D_{\rm eff}$ by  considering the overdamped limit of the combined particle and leg dynamics.
We introduce the 5 nondimensional scales
\begin{equation*}
  x \rightarrow L_x \tilde{x}, \,\, l_i - l_0 \rightarrow L \tilde{l}_i, \,\,  t \rightarrow \tau \tilde{t},  \,\,   v \rightarrow V \tilde{v},  \,\, m \rightarrow M \tilde{m}. 
\end{equation*}
Here $L = \sqrt{k_B T/k}$ is the characteristic length of leg fluctuations, while $L_x$ and $\tau$ are respectively the lengthscale and timescale for the long-time motion of $x$. 
The latter two scales are not determined \textit{a priori} by  intrinsic scales~\cite{fogelson2018enhanced,fogelson2019transport}, but rather are chosen large enough that coarse-graining will lead to diffusive dynamics~\cite{marbach2021nano}. 
Hence  $L_x = L/\epsilon$ where $\epsilon\ll 1$ is a small nondimensional number, and $\tau = L_x^2/D_0$, which corresponds to $\tau = \Gamma/k\epsilon^2$. 
Velocity fluctuations are fast compared to diffusive motion such that $V = L_x/\tau \epsilon$. Importantly, we specify the scale of mass by considering that the velocity auto-correlation time in the absence of recoil forces, $\tau_v = \tau_m = M/\Gamma $, is small compared to the observation time, $\tau_v = \tau_m = \epsilon^2 \tau$. This is the usual scaling to obtain \textit{overdamped}, or more generally long-time diffusive, dynamics~\cite{pavliotis2008multiscale}. 
Finally, we observe the system at sufficiently long times  that the remaining timescales are much shorter: $\gamma/k \sim \Gamma/k \sim \on^{-1} \sim \off^{-1}$, so that $\on \rightarrow \frac{\tau}{\epsilon^2} \tilde{q}_{\rm on}$ and similarly for $\off$.




We use these scalings to coarse-grain 
Eqns.~(\ref{eq:unbound}-\ref{eq:constitutiveBound}). 
Standard coarse-graining techniques~\cite{pavliotis2008multiscale,pavliotis2014stochastic,fogelson2018enhanced,lee2018modeling,marbach2021nano} (Supplementary \S2) 
show that the particle diffuses at long times with diffusion coefficient
\begin{equation}
D_{\rm eff}(m)   = \frac{k_B T}{\Gamma_{\rm eff}(m)} =  \sum_{n=0}^N p_n \frac{k_B T}{\Gamma_n(m)} \,,
\label{eq:Nlegs}
\end{equation}
where $p_n = \binom{N}{n} \frac{\on^n \off^{N-n}}{(\on + \off)^N}$ is the probability to have $n$ bonds and $\Gamma_n(m)$ are the effective friction coefficients for a state with $n$ bonds. The $\lbrace \Gamma_n \rbrace$ satisfy a linear system of equations 
that is reported in Eq.~(S2.23) and that depends on parameters $(m, N, k, \on, \off, \gamma, \Gamma)$. Importantly, Eq.~\eqref{eq:Nlegs} predicts up to order of magnitude changes on the effective diffusion $D_{\rm eff}$ depending on the specific microscopic parameter values. 

Let us analyze in detail a $N{=}1$-legged particle. The friction coefficients can be obtained analytically as
\begin{equation}
    \begin{split}
         \frac{\Gamma_0(m)}{\Gamma} & =\displaystyle 1 +  \frac{m q_{\rm on}}{\Gamma} \frac{\gamma_{\rm eff}
}{\gamma_{\rm eff} + \Gamma + m (q_{\rm on} + q_{\rm off}) } , \\
    \frac{\Gamma_1(m)}{\Gamma} & = 1 + \frac{\gamma_{\rm eff}}{\Gamma} \frac{\Gamma + m q_{\rm on}}{\Gamma + m (q_{\rm on} + q_{\rm off})}.
    \end{split}
    \label{eq:frictionsm}
\end{equation}
Here $\gamma_{\rm eff} = \gamma + k \left( \frac{1}{\off} + \frac{\gamma}{k} \frac{\on}{\off} \right)$ is the effective friction from the leg, including the leg's bare friction $\gamma$ and recoil forces from the tethered spring. 
The coefficients satisfy $\Gamma_0 \leq \Gamma_1$ as, when it is bound, the leg exerts additional recoil forces on the particle, as was observed in a variety of systems, from rubber~\cite{schallamach1963theory} to muscle friction~\cite{tawada1991protein} to virus motion on mucus~\cite{sakai2018unique}. We compare our analytic result for $D_{\rm eff}$ with direct stochastic simulations over a wide range of parameters and find excellent agreement (Figs.~\ref{fig:intro} and S1). Overall, the effective friction $\Gamma_{\rm eff}$ increases with mass, and therefore the particle's diffusion slows down with increased mass. 




Eq.~\eqref{eq:frictionsm} gives insight into what controls both the onset of the diffusion slow down, and the magnitude of the effect. 
Diffusion begins to decrease when the binding and unbinding times $\tau_{\rm on} = \on^{-1}, \tau_{\rm off}=\off^{-1}$ become comparable with the relaxation time of inertia, $\tau_m = m/\Gamma$. This is apparent in Fig.~\ref{fig:intro} where the transition between limit regimes occurs for $m \on/\Gamma \sim 1$. 
For shorter binding times, 
$\tau_{\rm on} \,{\sim}\, \tau_{\rm off}\ll\tau_m$, inertial effects matter, and the friction coefficients 
for $m \gg \Gamma/\on, \Gamma/\off$
converge to
\begin{equation}
    \frac{\Gamma^{m=\infty}_0}{\Gamma} =   \frac{\Gamma^{m=\infty}_1}{\Gamma} = 1 + p_1  \frac{\gamma_{\rm eff}}{\Gamma}.
    \label{eq:gammaMinf}
\end{equation}
The friction coefficients are the same, regardless of the state (bound or unbound) of the particle. This is coherent: since the particle has significant inertia, it does not have time to accelerate or decelerate to a different dynamical regime upon changing state. Binding and unbinding happen too rapidly for the particle to sense the difference. Eq.~\eqref{eq:gammaMinf} was also obtained in   Ref.~\cite{lee2018modeling} starting from the underdamped equations. 

Reciprocally, if binding timescales are long compared to inertial relaxation ($\tau_{\rm on} \sim \tau_{\rm off} \gg \tau_m$) we expect inertia to play a negligible role: the particle has time to accelerate and reach an overdamped limit motion before any further change of state occurs.  In this case the friction coefficients are
\begin{equation}
    \frac{\Gamma^{m=0}_0}{\Gamma} = 1, \quad  \frac{\Gamma^{m=0}_1}{\Gamma} = 1 + \frac{\gamma_{\rm eff}}{\Gamma}.
    \label{eq:gammaM0}
\end{equation}
Eq.~\eqref{eq:gammaM0} was also obtained in  Ref.~\cite{marbach2021nano}, starting with overdamped equations for the particle. 

We therefore find that the onset of inertial effects is governed by the ratio of timescales, $\tau_{\rm on} \sim \tau_{\rm off}$ compared to $\tau_m$. 
\textit{A posteriori}, it is natural that this onset is controlled by timescales, yet it was not obvious  \textit{which} of the diversity of timescales would matter.  
For example, the timescale for relaxation of leg fluctuations $k/\gamma$ does not control the occurrence of inertial slow-down. 

However, $k/\gamma$ does control the magnitude of the inertial slow-down, via $\gamma_{\rm eff}$.
%
The relative slow down between the underdamped and the overdamped regime is 
\begin{equation}
    \frac{D_{\rm eff}^{m = \infty}}{D_{\rm eff}^{m =0}} = \frac{1 + \frac{\gamma_{\rm eff}}{\Gamma}}{1 +  \frac{\gamma_{\rm eff}}{\Gamma} + p_0 p_1 \frac{\gamma_{\rm eff}^2}{\Gamma^2}}.
    \label{eq:slowdown}
\end{equation}
If the leg is very stiff ($k \gg \Gamma q_{\rm off}$, implying $\gamma_{\rm eff} \gg \Gamma$), then diffusion can be significantly slowed for massive particles, $D_{\rm eff}^{m = \infty} \ll D_{\rm eff}^{m = 0}$. Indeed, stiff legs greatly reduce motion in the bound state, $\Gamma_1 \gg \Gamma$. Since a heavy particle does not have the time to accelerate while its leg is unbound, we also have increased friction in the unbound state $\Gamma^{m=\infty}_0 \gg \Gamma$ and the particle's overall mobility is decreased, by up to orders of magnitude (as seen in Fig.~\ref{fig:intro}).
For an overdamped particle, on the contrary, even with a stiff leg, the particle can still move when it is unbound, as it has time to accelerate ($\Gamma^{m=0}_0 = \Gamma$), and its diffusion coefficient remains finite.

\begin{figure}[h!]
    \centering
    \includegraphics[width = 0.99\columnwidth]{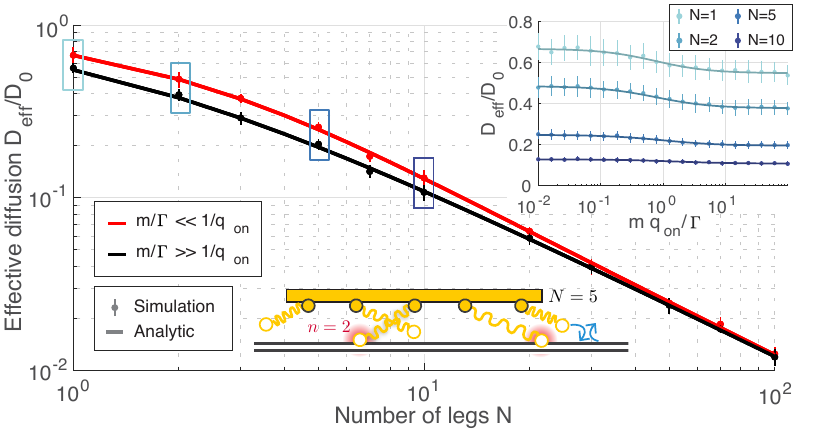}
    \caption{\textbf{Inertial slow down vanishes with numerous legs}. $D_{\rm eff}$ versus number of legs $N$, for large mass (black, $m \on / \Gamma = 100$) and small mass (red, $m \on / \Gamma = 0.01$), from stochastic simulations (markers) and from Eq.~\eqref{eq:Nlegs} (solid lines, solving numerically for $\lbrace \Gamma_n \rbrace$). (Top inset) $D_{\rm eff}$ versus inertia to binding timescale ratio for various $N$. (Bottom inset) Schematic of an $N$-legged particle including $n$ surface-bound legs.  Error bars are one standard deviation for 100 independent simulations, with $\gamma = 0.1 \Gamma$, $\off = 0.8 k/\Gamma$, and $\on = k/\Gamma$.}
    \label{fig:figparams}
\end{figure}

Let us now consider a  particle with many legs involved in the binding process, say $N\approx100-1000$, as in some DNA-coated colloids at low temperatures~\cite{xu2011subdiffusion,fan2021microscopic}. When the average number of bonds is large, $\overline{N}~=~\frac{\on}{\on + \off}~N~\gg~1$, the  $\lbrace\Gamma_n\rbrace$ can be approximated by the averaged value $\Gamma_{\overline{N}}$ (Supplementary \S2.3.4), yielding 
\begin{equation}
    \Gamma_{\rm eff}(m)  \stackrel[\overline{N} \gg 1]{}{\simeq}   \Gamma + \overline{N}\gamma_{\rm eff}.
    \label{eq:GNb}
\end{equation}
The diffusion coefficient no longer depends on the mass of the particle. Stochastic simulations, as well as numerical solutions of the linear system satisfied by the $\lbrace\Gamma_n\rbrace$, confirm this result: the diffusion  coefficient when $N$ is large converges to a value independent of mass (Fig.~\ref{fig:figparams}). Interestingly, the transition to the slowed-down diffusion regime  
occurs when $\tau_m/\tau_{\rm on} \propto N $ (Supplementary \S2.3.3). 

Why do inertial effects vanish with a large number of legs? For a heavy particle, the friction coefficients for each bond state are equal (as in Eq.~\eqref{eq:gammaMinf}):
$\Gamma^{m=\infty}_0 =  \Gamma^{m=\infty}_1 = ...  = \Gamma^{m=\infty}_N =  \Gamma + \overline{N} \gamma_{\rm eff}$ (Supplementary \S2.3.2), because the particle does not have time to accelerate between bond state changes, and hence is only sensitive to the average configuration. The difference is that now an average of $\overline{N}$ legs exerts extra recoil forces. For a very light particle, friction coefficients for each bond state are different, but their sum in Eq.~\eqref{eq:Nlegs} is dominated by the most likely state, which is  $\overline{N}$ when this average is large, leading to 
$\Gamma^{m=0}_{\rm eff} \approx \Gamma^{m=0}_{\overline{N}} =  \Gamma + \overline{N} \gamma_{\rm eff}$. Numerous legs can thus be thought of as self-averaging, and hence transitions between states do not matter as much as when there are a small number of bound legs.

\begin{figure}[h!]
    \centering
    \includegraphics[width = 0.95\columnwidth]{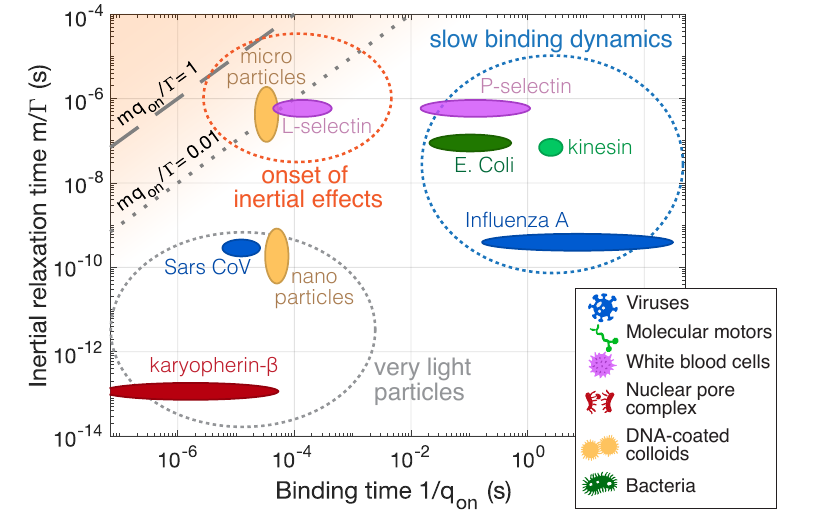}
    \caption{\textbf{Inertial effects in biophysical systems.} Ashby chart comparing the inertia and binding timescales for various systems. The height and width of the filled ellipses represent the range of values found in the literature (Supplementary~\S3). All data is from measurements at room temperature, except for DNA-coated colloids whose parameters are taken at their respective melting temperatures~\cite{fan2021microscopic}. 
    Large dashed circles represent global categories of systems. The shaded orange area  represents parameter values where inertial effects could be important.}
    \label{fig:figSystems}
\end{figure}

We now explore the possible emergence of such inertial effects in biological or biomimetic systems where a small average number of bound legs $\overline{N} \simeq 1$ is inherent, or can be achieved \textit{e.g.} with temperature control~\cite{fan2021microscopic}. 
To observe inertial effects, the binding times ($\tau_{\rm on} = \on^{-1}$, $\tau_{\rm off} = \off^{-1}$) have to be faster than the inertial relaxation time $\tau_m = m/\Gamma$. As typical adhesive systems have 
$\off \lesssim 10 \on$, we focus 
on $\on$. 
We report in Fig.~\ref{fig:figSystems} the orders of magnitude for the momentum relaxation time $\tau_m = m/\Gamma$ and binding times $\tau_{\rm on} =\on^{-1}$ for a variety of particles with ligand-receptor contacts using data available in the literature~\cite{sauter1989hemagglutinins,reiter2019force,wang2015crystallization,xu2011subdiffusion,zhang2018predicting,fan2021microscopic,park2008dna,fakih2017gold,hurst2006maximizing,ting1993volume,shao1998static,dwir2003avidity,schwarz2004selectin,fritz1998force,mehta1998affinity,alon1995lifetime,alon1995lifetime,alon1997kinetics,bakshi2012superresolution,chen2001selectin,miller2006mechanical,yakovenko2015inactive,sauer2018binding,sauer2016catch,neidhardt1990physiology,korn2009stochastic,gibbons2001dynamical,wu2009intracellular,aramburu2017floppy,radu1995identification,aramburu2017floppy,milles2015plasticity,tu2013large,harris2006influenza,sieben2012influenza,muller2019mobility,laue2021morphometry,yang2020molecular,laue2021morphometry,yang2020molecular,walls2020structure} (recapitulated in Supplementary \S3.2).

Numerous ligand-receptor systems have binding kinetics that are 
too slow to observe inertial effects, with $\on\lesssim100~\mathrm{s}^{-1}$ (blue dashed circle in Fig.~\ref{fig:figSystems}) -- \textit{e.g.} spike proteins on the Influenza A virus~\cite{sauter1989hemagglutinins,reiter2019force}, molecular motors transporting cargos~\cite{vale1996direct}, and pili adhesion of Escherischia Coli~\cite{yakovenko2015inactive,sauer2018binding,sauer2016catch}. 

Other systems have fast binding kinetics ($\on\gtrsim10^{4}~\mathrm{s}^{-1}$), but not fast enough to incur inertial effects on the lighter systems they are connected to (gray dashed circle in Fig.~\ref{fig:figSystems}). These often correspond to smaller particles, and since $ m/\Gamma \propto R^2$, this decreases the maximum binding timescale required to observe inertial effects. Examples include Sars CoV 1 and 2 viruses~\cite{yang2020molecular,walls2020structure},  DNA-coated nanocolloids~\cite{park2008dna} and protein transporters in the nuclear pore complex~\cite{radu1995identification}. %

Inertial effects may occur for two systems with a specific combination of large particles and fast binding kinetics (orange dashed circle in Fig.~\ref{fig:figSystems}): (a) micron-sized DNA-coated colloids near their melting temperature 
and (b) white blood cells with adhesion mediated by L-selectin. For both systems, typical existing experimental designs possess an inertia to binding timescale ratio $m \on /\Gamma \simeq 10^{-3} - 10^{-1}$, which is close to the range where we predict inertial slow down.

DNA-coated colloids offer a promising route to probe such inertial slow down,  as they may be finely tuned by changing their size, coating density, ligand length, material composition, \textit{etc} and to observe diffusion over several degrees~\cite{wang2015crystallization,marbach2021nano}.
We speculate that inertial effects could be observed by solving the challenge of building such colloids with different cores~\cite{sacanna2007generic,fakih2017gold} but keeping the surface DNA coatings the same: using a gold~\cite{xu2011subdiffusion} or a polystyrene~\cite{fan2021microscopic} core to make a heavy or a light particle respectively. 

To maximize inertial effects one must increase both $m \on / \Gamma$ -- to reach the transition between the non-inertial and inertial regimes -- as well as $\gamma_{\rm eff}/\Gamma$ -- to increase the magnitude of the slow down, see Eq.~\eqref{eq:slowdown}. The latter ratio, in DNA-coated colloids, is typically dominated by the term $k/\Gamma \off$. This points to the dual role of certain parameters. For example, while increasing the particle radius increases  $m \on / \Gamma$ (since $m \propto R^3$ and $\Gamma \sim R$), it decreases  $k/\Gamma \off$, and hence an optimal particle radius is needed. Furthermore, while shorter polymer lengths increase stiffness $k$, they also increase hydrodynamic friction $\Gamma$ through lubrication effects~\cite{goldman1967slow,sprinkle2020driven} and hence an optimal polymer length is also required. We elaborate in detail on the role of various experimental parameters in Supplementary \S4 and we provide a predictive tool for the rapid exploration of different material designs~\footnote{Our tool is available through GitHub: https://github.com/smarbach/DNACoatedColloidsInteractions}.


We estimate the change in diffusion coefficient as a function of temperature that using different cores would induce (Supplementary \S4), computing $D_{\rm eff}$ using Eq.~\eqref{eq:Nlegs}. We compute the temperature dependence of $N$ and $\off$ using a mean-field model that has been validated experimentally~\cite{fan2021microscopic,milner1988theory,rubinstein2003polymer,santalucia1998unified,chen2012ionic}, observing that $\on, \gamma,\Gamma$ do not change significantly with temperature~\cite{zhang2018predicting}. 
The difference between the diffusion coefficients of gold and polystyrene colloids is maximal at intermediate temperatures where colloids form only a few bonds ($\overline{N} = 1 - 10$) with the surface (Fig.~S3-S6). 
Drawing parameters from existing particle systems, we predict the difference in diffusion coefficients to be $1-2\%$ (Fig.~S3). 
While this is a small effect at the single particle level, mass discrepancies between numerous particles, and hence diffusion coefficient discrepancies, could accumulate to impact collective properties such as nucleation, annealing (as was seen in a different context~\cite{ramananarivo2019Activitycontrolled}) or trigger mass-dependent phase separations. 
Fine tuning particle coatings, \textit{e.g.} reducing the ligand length with commercially available ligands, could increase the difference in single particle diffusion coefficients to $6-7\%$ (Fig.~S4-S5), which is well within experimental accuracy~\cite{marbach2021nano}. Exploiting further advanced experimental conditions such as changing the solvent~\cite{thompson2006viscosity,nakano2016structural} increases the discrepancy to $10-20\%$ (Fig.~S6). 



In summary, our model predicts that inertia could modify the diffusion coefficient of particles in fluids with ligand-receptor contacts, inducing a diffusion slow-down with increased inertia. 
The onset of the slow-down occurs when the binding timescale  $\tau_{\rm on}=\on^{-1}$ is faster than the timescale for the inertial relaxation, which is $\tau_m = m/\Gamma$ in our model, a lower bound on the actual timescale since momentum decays algebraically in most fluid systems~\cite{bian2016111,balboa2013stokes}. 
The magnitude of the inertial slow-down is 
increased with stiff ligands and fewer bound legs. Improvements to our model could include, among other things, fluid memory kernels to investigate the algebraic decay of momentum~\cite{bian2016111} or ligand density inhomogeneities to probe subdiffusive dynamics that are observed at low temperatures~\cite{xu2011subdiffusion}. As the main principles inducing mass-dependent dynamics are essential to the account of ligand-receptors, namely binding and unbinding and altered motion when bound, it is reasonable to assume that mass-dependent diffusion should persist in any ligand-receptor model. 
We predict the diffusion slow down could be probed experimentally by fine-tuning DNA-coated colloids. 

Our analysis thus provides a key principle to investigate the onset of inertial effects in other micronscale particles in liquids. When there exists a physical timescale in the system that is fast, and comparable to the relaxation of inertia, inertial effects \textit{could} arise. This criterion is repeatedly observed in other contexts~\cite{schmidt2003hydrodynamic,balboa2013stokes,bian2016111,li2013brownian,daddi2016long,trulsson2012transition,boyer2011unifying,demery2015mean,lowen2020inertial,dai2020phase,lindner1999inertia}.
However, in general inertial effects do not necessarily imply that the diffusion coefficient depends on mass. 
For example, when a particle has an inertial relaxation time comparable to the relaxation time of the solvent, then the particle's velocity autocorrelation function decays algebraically instead of exponentially, but its diffusion coefficient remains independent of inertia~\cite{bian2016111}. Hence, an overdamped, equilibrium system where single-particle diffusion depends on mass remains surprising.

Targetted experiments, especially on particles with ligand-receptor contacts, could identify other inertial effects beyond diffusion slow down. 
For DNA-coated colloids, one could envision that such inertia-modified dynamics could also impact collective properties such as crystallographic alignment into self-assembled structures~\cite{jenkins2014hydrodynamics,holmes2016stochastic}. Understanding the dynamics of such complex micronscale particles is a key step to pave the way towards controlled design at the microscale, \textit{e.g.} to improve synthesis of materials with advanced optical properties~\cite{color2015PNAS,he2020colloidal}.


\begin{acknowledgments}
The authors thank Brennan Sprinkle for sharing his code to calculate lubrication forces, and Aleksandar Donev for fruitful discussions.
S.M. received funding from the European Union’s Horizon 2020 research and innovation programme under the Marie Skłodowska-Curie grant agreement 839225, MolecularControl.
All authors were supported in part by the MRSEC Program of the National Science Foundation under Award Number DMR-1420073. M.H.-C. was partially supported by the US Department of Energy under Award No. DE-SC0012296, by grant NSF-DMS-2111163, and acknowledges support from the Alfred P. Sloan Foundation.
\end{acknowledgments}




%


\end{document}